\documentclass[12pt,english,1p]{elsarticle}
\usepackage[T1]{fontenc}
\usepackage[latin9]{inputenc}
\usepackage{array}
\usepackage{float}
\usepackage{amsbsy}
\usepackage{graphicx}
\usepackage{setspace}
\usepackage{esint}

\makeatletter
\providecommand{\tabularnewline}{\\}

\@ifundefined{showcaptionsetup}{}{%
 \PassOptionsToPackage{caption=false}{subfig}}
\usepackage{subfig}
\makeatother
\journal{Wave Motion}
\usepackage{babel}

\begin{document}
\begin{frontmatter}
\title{Modal analysis of the scattering coefficients of an open cavity in
a waveguide }

\author[uwa]{Yuhui Tong}

\author[uwa]{Jie Pan\corref{cor1}}
\cortext[cor1]{Corresponding author}
\ead{jie.pan@uwa.edu.au}

\address[uwa]{School of Mechanical and Chemical Engineering, The University of
Western Australia, Crawley, WA 6009, Australia}

\begin{abstract}
The characteristics of an acoustic scatterer are often described by
scattering coefficients. The understanding of the mechanisms involved
in the frequency dependent features of the coefficients has been a
challenge task, owing to the complicated coupling between the waves
in open space and the modes inside the finite scatterer. In this paper,
a frequency-dependent modal description of the scattering coefficient
is utilized to study the modal properties of the scatterer. The important
role that eigenmodes play in defining the features of the scattering
coefficients is revealed via an expansion of the coefficients by the
eigenmodes. The results show the local extrema of the scattering coefficients
can be attributed to the constructive/destructive interference of
resonant and non-resonant modes. In particular, an approximated equation,
which is equivalent to the standard Fano formula, is obtained to describe
the sharp anti-symmetric Fano characteristics of the scattering coefficients.
The special cases where scattering is dominated by a single resonance
eigenmode, corresponding to the ``resonance transmission'', are
also illustrated.\end{abstract}
\end{frontmatter}

\section{Introduction}

The scattering coefficients of acoustic scatterers are important parameters
as they are used to relate incident and scattered waves. Traditionally,
sound scattering in a duct by a muffler is described by one-dimensional
(1D) transmission line theory, and explained by the mismatch of the
specific impedance at the inlet and outlet of themuffler \cite{Kinsler2000}.
To accommodate cross modes in a duct with scatterers, the wave matching
technique \cite{selamet1999acoustic,kirby2005point} is used. The
finite element and boundary element methods, are used to determine
the scattering coefficients when the geometries of the ducts and scatterers
are complicated \cite{graf2013determination,wang2014impedance}. While
these wave matching and numerical methods are capable of producing
accurate coefficients, they are less useful in directly delivering
the physical insight into the peaks and valleys in the coefficient
curves.

An alternative approach, which is motivated by the obervation of trapped
and quasi-normal modes inside the scatterers, is to describe the scattering
in terms of the coupling between the waves in the ducts and modes
inside the scatterer. This was inspired by the early work of Flax
\emph{et al.} \cite{flax1978theory}, where the sound scattering from
submerged elastic bodies is affected by various kinds of interference
between the resonance scattering at the eigenfrequencies of the vibration
of the body and the rigid-body scattering. Recently, such approach
has been used to explain the peaks and valleys in the transmission
loss curve of an expansion chamber subject to an incident plane-wave
in a 1D duct \cite{pan2015QuasiNormalModes}. It was demonstrated
that the characteristics (complex eignevalues and mode shape functions)
of the frequency-dependent quasi-normal modes of the expansion chamber
\cite{kergomard2006resonance} allow the correct expansion of the
sound pressure in the chamber. The coupling of the quasi-normal modes
with the incident and transmitted sound waves sheds some light on
the transmission loss. For example, the minimum values in the transmission
loss occur when the frequency of the incident wave equals the real
part of the eigenfrequency of the quasi-normal mode. On the other
hand, the maximum transmission loss appears at those frequencies where
the superimposed contribution from the participated modal factors
is at minimum. However, the extension of these previous works to two-dimensional
(2D) or three-dimensional (3D) expansion chambers and modelling of
the coupling between the quasi-modes in the chambers and the incident
and transmitted waves including cross modes, is not straight forward.
Furthermore, because of the involvement of cross mode components in
the 2D and 3D configurations, extra coupling mechanisms due to participation
of the cross modes may lead to a more complete understanding of the
characteristics of the scattering coefficients.

Most recently, Maksimov \emph{et al}. \cite{maksimov2015coupled}
and Lyapina \emph{et al}. \cite{lyapina2015bound} looked at the aforementioned
scattering problem from a more fundamental view of the acoustical
properties of an open cavity. An open cavity is a finite acoustical
space with defined boundary conditions at its internal wall (\emph{e.g.,
a} rigid-wall condition) and some boundary areas open to infinite
space(s), such as semi-infinite waveguides. Because of the energy
exchange between the sound fields inside the open-cavity and the infinite
space(s), the sound field inside the cavity is characterized by the
non-Hermitian Hamiltonian. They used the acoustic coupled mode theory
for the calculation of the frequency-independent eigensolutions of
the open cavity and revealed a mechanism resulting in acoustic trapped
modes in the cavity, namely Friedrich\textendash Wintgen two-mode
full destructive interference. Although being used to explain the
occurrence of Fano resonance, the nonorthorgonal properties and incompleteness
of the frequency-independent eigensolutions of the open cavity make
the direct modal interpretation of the transmission coefficient (which
is a forced scattering problem at the frequency of an incident wave)
a difficult task. Xiong \emph{et al}. \cite{xiong2016fano}, on the
other hand, derived the scattering matrix of the open cavities due
to incident waves from one of the waveguide connected to the cavity.
Frequency-dependent eigensolutions of the effective Hamiltonian matrix
(including the interaction between cavity and connected waveguides)
for the sound field in the cavity were used to describe the scattering
coefficients and to explain the links between the eigenmodes and a
trapped mode and the corresponding transmission zero. Their contribution
makes possible an analysis of the scattering coefficients of the open
cavities by using the frequency-dependent eigensolutions of the cavity. 

In this paper, the method for frequency-dependent eigenmodes and the
scattering matrix developed by Xiong \emph{et al}. \cite{xiong2016fano}
is adopted. Instead of focusing on the Fano resonance induced by trapped
modes, it is used to calculate and explain the general scattering
features of the open cavities connected with waveguides (\emph{e.g.,}
conventional muffler configuration) and the roles played by eigenmodes
in determining the frequency-dependent features of the scattering
coefficients. Through numerical studies, it is revealed that extrema
in scattering coefficients are \emph{generally} a result of interference
between eigenmodes, rather than the contribution from single resonant
mode, as is traditionally assumed. The Fano resonance induced by highly
localized modes (quasi-trapped modes), as observed by Hein \emph{et al.} \cite{hein2012trapped},
is also revisited in terms of frequency-dependent eigenmodes. Finally,
some remarks are made to clarify the usage of frequency-dependent
and frequency-independent modes in conducting modal analysis in scattering
problems. 


\section{Modal description of the scattering coefficients\label{sec:Theory}}

This paper considers the scattering problem in a 2D acoustic scatterer,
comprising a cavity connected by $N$ uniform ducts. Omitting the
time-dependence term, $e^{-i\omega t}$, the sound pressure field
is governed by Helmholtz equation.

\begin{equation}
(\frac{\partial^{2}}{\partial x^{2}}+\frac{\partial^{2}}{\partial y^{2}}+k^{2})p(x,y)=0,\label{eq:helmhotlzequation}
\end{equation}
and corresponding boundary conditions, where $k$ is the wavenumber.
Since $k=\omega/c_{0}$ (where $c_{0}$ is the speed of sound), $k$
will be used hereafter to represent source frequency, \emph{i.e.},
whenever ``frequency'' is mentioned, it refers to the wavenumber.
For the sake of simplicity, the rigid-wall boundary condition is assumed
for the cavity and ducts. Although details of the coupled mode theory
\cite{maksimov2015coupled,lyapina2015bound,xiong2016fano} have already
been published, this paper provides a brief description of the derivation
using coupled mode theory for the scattering coefficients of the scatterer
in terms of frequency-dependent eigenmodes, for the convenience of
the reader and to make the paper self-contained.

The first step is to express the sound pressure in terms of a local
basis in different regions. The geometry is partitioned into $(N+1)$
regions: a closed cavity $\Omega_{c}$ and \emph{N} semi-infinite
ducts $\Omega_{n}$ ($n=1,2,...,N$). The pressure field in the $n^{th}$
duct is expanded into duct modes when taking a local coordinate $(x_{n},y_{n})$
with $x-$ and $y-$axes that are, respectively, perpendicular and
vertical to the $n^{th}$ duct-cavity interface:

\begin{equation}
p(x_{n},y_{n})=\sum_{p}(a_{n,p}e^{-i\kappa_{n,p}x_{n}}+b_{n,p}e^{-i\kappa_{n,p}x_{n}})\chi_{n,p}(y_{n}),\label{eq:ductmodeexpansion}
\end{equation}
where $a_{n,p}$ and $b_{n,p}$ are respectively, the amplitude for
the $p^{th}$ \emph{incident} and \emph{scattered} modes. The transverse
functions $\chi_{n,p}(y)$ are given by $\chi_{n,p}(y_{n})=\sqrt{\frac{2-\delta_{0,p}}{d_{n}}}\cos(\frac{p\pi}{d_{n}}y_{n}),$
where $d_{n}$ is the width of the $n^{th}$ duct. Finally, $\kappa_{n,p}$
is the axial wavenumber such that, $\kappa_{n,p}=\sqrt{k^{2}-(p\pi/d_{n})^{2}},\ (p=0,1,2,....)$ 

The pressure field in the cavity may be expanded by closed cavity
modes such that

\begin{equation}
p(x,y)=\sum_{\mu}g_{\mu}\psi_{\mu}(x,y).\label{eq:rigidcavityexpansion}
\end{equation}
where $\psi_{\mu}(x,y)$ is the \emph{closed} cavity mode satisfying

\begin{equation}
(\frac{\partial^{2}}{\partial x^{2}}+\frac{\partial^{2}}{\partial y^{2}}+k_{\mu}^{2})\psi_{\mu}(x,y)=0,\label{eq:orthogonalbasis}
\end{equation}
with Numann (rigid) boundary conditions at the duct-cavity interfaces
and at other boundary surface of the cavity, where $k_{\mu}$ is the
eigenvalue. 

\begin{figure}
\begin{centering}
\includegraphics[scale=0.35]{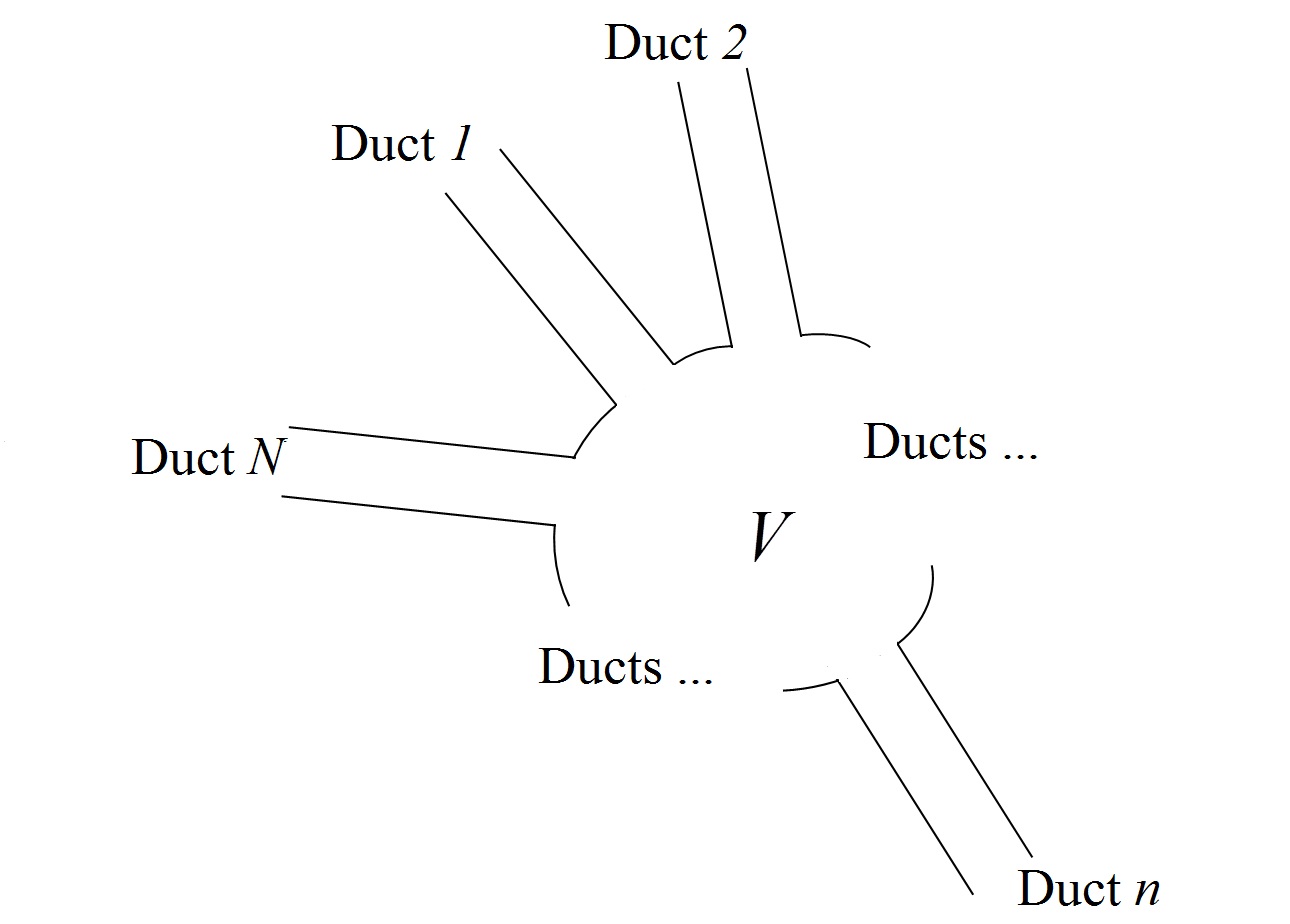}
\par\end{centering}

\caption{An open cavity connected with \emph{N} uniform rigid-wall ducts.}
\end{figure}

Although $\psi_{\mu}(x,y)$ does not satisfy the continuity condition
for particle velocity at the cavity-duct interfaces, it is worth emphasizing
that the expansion in Eq. (\ref{eq:rigidcavityexpansion}) is valid
as the matching conditions for particle velocity must not be imposed
directly, which is a consequence of the fact that the convergence
in Hilbert space does not imply point-wise convergence at the boundary
\cite{viviescas2003openopticalcavities,jackson1999classical}. 

In a third step, a key equation is derived to determine the coefficient
$g_{\mu}$ for the cavity upon incident waves by applying the continuity
condition of pressure and velocity in a limiting sense infinitesimally
close to the boundary:

\begin{equation}
\left(\boldsymbol{D}-k^{2}\boldsymbol{I}\right)\boldsymbol{g}=-2i\sum_{n=1}^{N}\sum_{p=1}^{+\infty}\kappa_{n,p}\boldsymbol{v}_{n,p}a_{n,p},\label{eq:keyequation_cavitymodes}
\end{equation}
where $\boldsymbol{I}$ is the identity matrix, $\boldsymbol{g}$
is an unknown column vector containing the coefficients $g_{\mu}$.
The column vector $\boldsymbol{v}_{n,p}$ contains the coupling constant
$v_{n,p}^{\mu}$ between the rigid cavity mode $\psi_{\mu}$ and the
duct mode $\chi_{n,p}$ \cite{maksimov2015coupled,pichugin2001openbiliiards},\emph{
i.e.}, $v_{n,p}^{\mu}=\int\psi_{\mu}\chi_{n,p}d\tau_{n},$ where the
integral is evaluated over the interfaces between the cavity and the
$n^{th}$ duct, and $\tau_{n}$ is the corresponding surface element.
The matrix $\boldsymbol{D}$ has the following form:

\begin{equation}
\boldsymbol{D}=\boldsymbol{k}_{c}^{2}-i\sum_{n=1}^{N}\sum_{p=1}^{+\infty}\kappa_{n,p}\boldsymbol{v}_{n,p}\boldsymbol{v}_{n,p}^{\dagger},\label{eq:effectivehamiltonian}
\end{equation}
where $\boldsymbol{k}_{c}^{2}$ is a diagonal matrix containing the
squared eigenfrequencies $k_{\mu}^{2}$, and symbol $\dagger$ denotes
the conjugate transpose. It then follows that the scattered wave $b_{n,p}$
(the $p^{th}$ outgoing duct mode in the $n^{th}$ duct) is found
as 

\begin{equation}
b_{n,p}=-a_{n,p}+\boldsymbol{v}_{n,p}^{\dagger}\boldsymbol{g}.\label{eq:scatteredwave}
\end{equation}

The scattering coefficients $s_{n,n',p,p'}$ are defined by $b_{n,p}=\sum_{n',p'}s_{n,n',p,p'}a_{n',p'}$.
Substituting the solution of Eq. (\ref{eq:keyequation_cavitymodes})
into Eq. (\ref{eq:scatteredwave}) yields

\begin{equation}
s_{n,n',p,p'}(k)=-\delta_{n,n'}\delta_{p,p'}-2i\boldsymbol{v}_{n,p}^{\dagger}\frac{1}{\left(\boldsymbol{D}-k^{2}\boldsymbol{I}\right)}\kappa_{n',p'}\boldsymbol{v}_{n',p'},\label{eq:acousticscatteringformula}
\end{equation}
which relates the scattered wave $b_{n,p}$ to the incident wave $a_{n',p'}$.

.

In a fourth step, an eigenvalue problem (EVP) of $\boldsymbol{D}$
subject to the source frequency $k$ is solved as 
\begin{equation}
\boldsymbol{D}(k)\boldsymbol{G}_{\mu}=K_{\mu}^{2}\boldsymbol{G}_{\mu},\label{eq:EVP2}
\end{equation}
 where $K_{\mu}^{2}$ and $\boldsymbol{G}_{\mu}$ are respectively
the frequency-dependent eigenvalue and eigenvector. Since $\boldsymbol{D}(k)$
is a non-Hermitian matrix, its eigenvectors satisfy the bi-orthogonal
relation (see Appendix),

\begin{equation}
\boldsymbol{G}_{\mu'}^{T}\boldsymbol{G}_{\mu}=\delta_{\mu',\mu},\label{eq:bi-orthogonalrelation}
\end{equation}
where the superscript $T$ denotes a transpose. Alternatively, Eq.
(\ref{eq:bi-orthogonalrelation}) may be written as

\begin{equation}
\iint_{\Omega_{c}}\Psi_{\mu'}(x,y)\Psi_{\mu}(x,y)ds_{c}=\delta_{\mu,\mu'},
\end{equation}
where $\Psi_{\mu}(x,y)$ is the mode-shape function corresponding
to $\boldsymbol{G}_{\mu}$. Note that the integral is evaluated over
$\Omega_{c}$, the domain occupied by the cavity,. 

Finally, expanding $\boldsymbol{g}$ in Eq. (\ref{eq:keyequation_cavitymodes})
in terms of eigenvectors of $\boldsymbol{D}(k)$,\emph{ i.e.}, $\boldsymbol{g}=\sum_{\mu}c_{\mu}\boldsymbol{G}_{\mu},$
and substituting it into Eq. (\ref{eq:scatteredwave}) yields the
expression for scattering coefficients in terms of eigenmodes

\begin{equation}
s_{n,n',p,p'}=-\delta_{n,n'}\delta_{p,p'}+\sum_{\mu}\frac{-2i\kappa_{n',p'}H_{\mu,n',p'}H_{n,p,\mu}}{K_{\mu}^{2}-k^{2}},\label{eq:new scattering formula}
\end{equation}
where $H_{\mu,n',p'}=\boldsymbol{G}_{\mu}^{T}\boldsymbol{v}_{n',p'}$
and $H_{n,p,\mu}=\boldsymbol{v}_{n,p}^{\dagger}\boldsymbol{G}_{\mu}$. 

Compared with Eq. (\ref{eq:acousticscatteringformula}), Eq. (\ref{eq:new scattering formula})
expresses the scattering coefficients explicitly as a superposition
of individual eigenmode components. The first term on the right-hand
side of Eq. (\ref{eq:new scattering formula})comes from the direct
reflection, while each term in the summation corresponds to scattering
by eigenmode $\boldsymbol{G}_{\mu}$. The term $H_{\mu,n',p'}$ measures
the coupling between eigenmode $\boldsymbol{G}_{\mu}$ and the incident
duct mode $(n',p')$. Similarly, $H_{n,p,\mu}$ provides a measurement
of the coupling between eigenmode $\boldsymbol{G}_{\mu}$ and the
scattered duct mode $(n,p)$.


\section{Results and discussions}

The aim of this section is to demonstrate the role that the eigenmodes
of the acoustic scatterer play in sound scattering, which is a question
originally put forward by \cite{yu2016acoustic} in their analysis
of wave propagation in continuous right-angled bends. This question
can now be answered mathematically using Eq. (\ref{eq:new scattering formula}),
by showing that scattering coefficients can indeed be expressed in
terms of the eigenmodes of an acoustic scatterer. Numerical results
will be used to support this answer.

In Eq. (\ref{eq:new scattering formula}), the summation terms are
characterized by the inverse of $(K_{\mu}^{2}-k^{2})$. One might
therefore assume that $s_{n,n',p,p'}$ is dominated by each eigenmodes
within each frequency range and that $|s_{n,n',p,p'}|$ achieves extreme
values. The numerical study, however, will show that the local extrema
of the scattering coefficients, are generally the result of constructive
or destructive interference of eigenmodes, rather than by dominating
resonance alone as one would expect intuitively. In the following
parts of this section, the typical interference of eigenmodes will
be covered first in Sec. 3.1. Then Sec. 3.2 will discuss a special
narrow band interference known as Fano Resonance. Section 3.3 will
be devoted to the phenomenon of resonance transmission. 

\begin{figure}
\hfill{}\subfloat[]{\centering{}\includegraphics[scale=1.5]{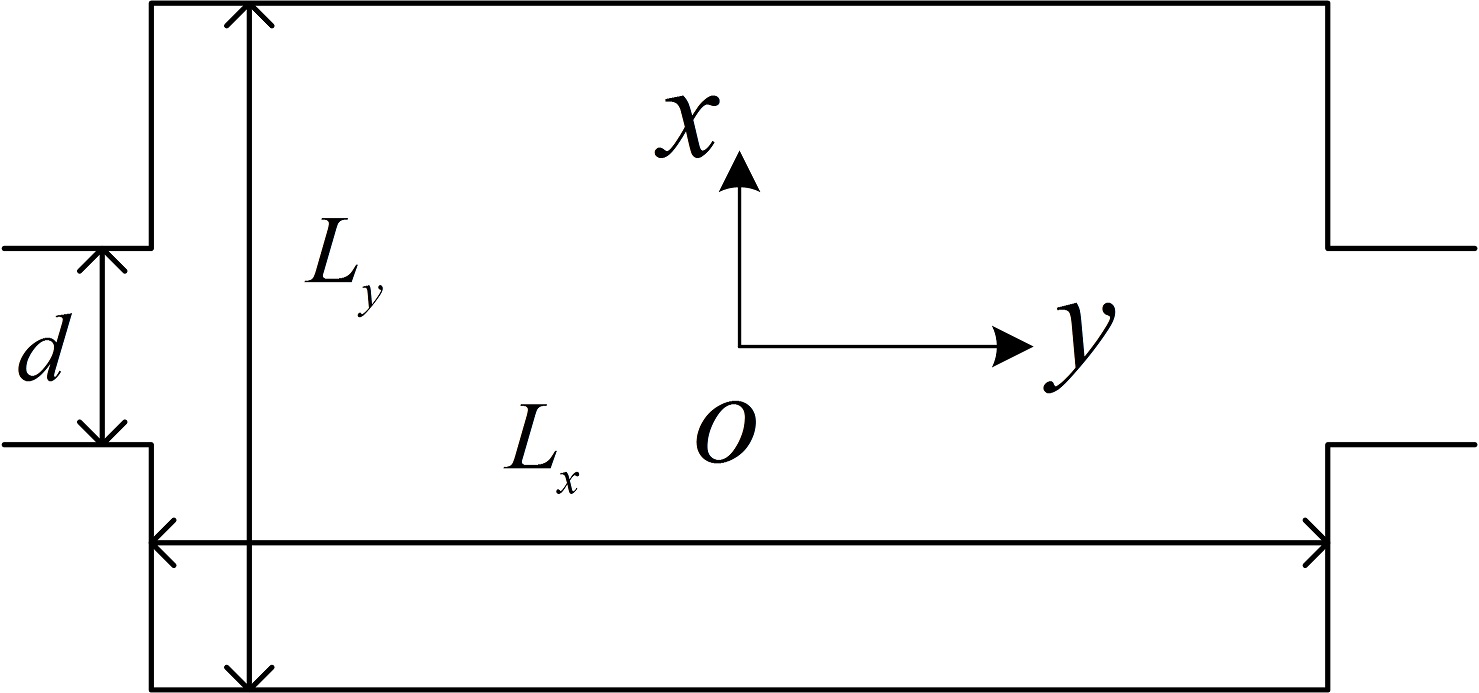}}\hfill{}\subfloat[]{\centering{}\includegraphics[scale=1.5]{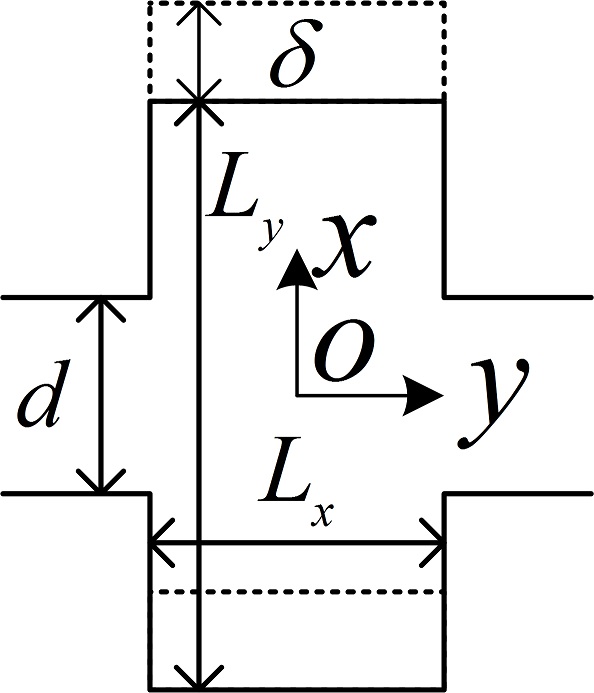}}\hfill{}

\caption{\label{fig:Geometry of resonators}Geometries of (a) long cavity (b)
short cavity.}
\end{figure}

It is worth noting that, although the theory proposed in Sec. \ref{sec:Theory}
is valid for scatterers of arbitrary-shape and with multiple connecting
ducts, for the sake of simplicity, the numerical study will be restricted
to scatterers consisting a rectangular cavity connected by two uniform
ducts, with the geometries illustrated in Fig. \ref{fig:Geometry of resonators}.
The scatterer with two connecting ducts is the simplest yet sufficiently
representative of its kind, whereas a rectangular cavity enables an
analytical solution of the closed cavity modes $\psi_{\mu}(x,y)$
; otherwise, the complexity of mode classification and subsequent
analysis may mask important physical phenomena.

In the following context, symbols L and R are used to denote ducts
on the left- and right-hand sides, respectively. The scattering coefficients,
owing to the symmetry of the systems, are reduced to the transmission
and reflection coefficients, $T_{p,p'}$ and $R_{p,p'}$, respectively.
For simplicity, only $T_{p,p'}$ is examined here. Using Eq. (\ref{eq:new scattering formula})
and modal decomposition by each eigenmode $\mu$, $T_{p,p'}$ is expressed
as

\begin{equation}
T_{p,p'}=\sum_{\mu}T_{p,p',\mu},\label{eq:Tdecomposition}
\end{equation}
where 
\begin{equation}
T_{p,p',\mu}=\frac{-2i\kappa_{n',p'}H_{\mu,n',p'}H_{n,p,\mu}}{K_{\mu}^{2}-k^{2}}.\label{eq:MPT}
\end{equation}
and $\mu$ is the index of the $\mu^{th}$ bi-orthogonal eigenmodes.
For a rectangular cavity where variable separation is possible, $\mu$
is represented by a pair of non-negative values $(m,n)$. The analysis
of $R_{p,p'}$ is omitted as it can be conducted following the same
procedures.

\subsection{Typical case: interference of eigenmodes}

The first to be considered is a short expansion cavity, depicted by
Fig. 2(b), which supports a typical case of interference for eigenmodes.
The system parameters are $d=1$ m, $L_{x}=1$ m and $L_{y}=3$ m,
and an adjustable vertical offset $\delta$ is used to dictate the
geometrical symmetry about the $x-$axis ($\delta=0$ is considered
here). Since the length of the cavity, $L_{y}$, is comparable with
its width, $L_{x}$, one can anticipate that the 2D modes with non-uniform
transverse functions may play an important role in sound scattering.

Figure 3(a) presents the amplitude of $|T_{0,0}|$ versus frequency.
Several peaks are observed at $k=0\ \mathrm{m^{-1}}$ and $k=2.94\ \mathrm{m^{-1}}$
and several dips at $k=1.68\ \mathrm{m^{-1}}$ and $k=3.42\ \mathrm{m^{-1}}$.
Using Eq. (\ref{eq:Tdecomposition}), one can decompose $|T_{0,0}|$
into a contribution by individual eigenmodes. Due to the transversal
symmetry of the system, only transversally symmetric modes can participate
in the scattering of the plane wave, \emph{i.e.}, only the (0,0),
(0,2) and (1,0) modes need to be taken into account for $k<3.5\ \mathrm{m^{-1}}$.
Figure. 3(d) and (e) show $K_{\mu}$, which are the eigenvalues of
the corresponding modes. It is noted that $Re(K_{\mu})$ ($\mu=(0,0),(0,2)\ \mathrm{and}\ (1,0)$
) varies slowly with \emph{k} except for $Re(K_{(0,0)})$, while $Im(K_{\mu})$
always start from 0 when $k=0\ \mathrm{m^{-1}}$ and decreases rapidly
as \emph{k} increases. Figure 3(d) shows the components of $T_{0,0}$,
\emph{i.e.}, $T_{0,0,\mu}$ ($\mu=(0,0),(0,2),(1,0)$), as functions
of \emph{k}. When \emph{k} approaches $Re(K_{\mu})$, the modal transmission
coefficient of the $\mu^{th}$ mode experiences resonance, leading
to the resonance peak of each $|T_{0,0,\mu}|$, \emph{e.g.}, at $k=0\ \mathrm{m^{-1}}$
for $\mu=(0,0)$, at $k=2.18\ \mathrm{m^{-1}}$ for $\mu=(0,2)$ and
at $k=3.42\ \mathrm{m^{-1}}$ for $\mu=(1,0)$. The width of each
resonance peak, on the other hand, is dictated by $Im(K_{\mu})$,
\emph{e.g.}, $|T_{0,0,(0,2)}|$ has a wide peak as $Im(K_{(0,2)})$
grows large, while the peak of $|T_{0,0,(1,0)}|$ is relatively narrow
as $Im(K_{(1,0)})$ grows small. 

A comparison between Fig 3(a) and Fig. 3(b) indicates that there are
no direct relations between the peaks of $|T_{0,0,\mu}|$ and the
extrema of $|T_{0,0}|$ (except for the first peak at $k=0\ \mathrm{m^{-1}}$,
which will be discussed in Sec. 3.3). Instead, the local minimum in
$|T_{0,0}|$ at $k=1.68\ \mathrm{m^{-1}}$ is due to the destructive
interference between the (0,2), (0,0) and (1,0) modes. The local maximum
at $k=2.9\ \mathrm{m^{-1}}$ arises from the constructive interference
between (0,2), (1,0), and (0,0) eigenmodes, whereas the destructive
interference between (0,2) and (1,0) gives rise to the local minimum
of $|T_{0,0}|$ at $k=3.42\ \mathrm{m^{-1}}$.

\begin{figure}
\begin{centering}
\includegraphics[scale=1]{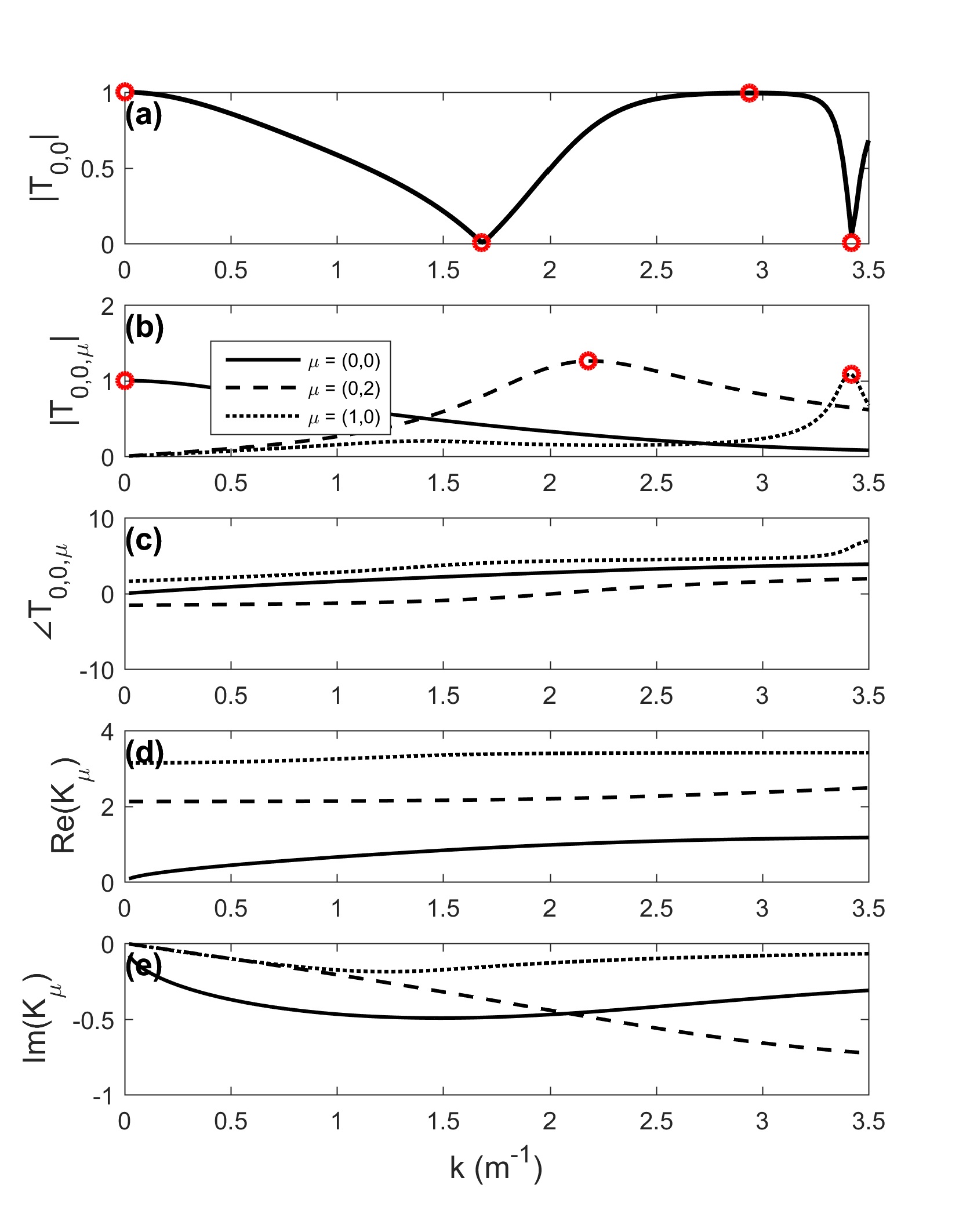}
\par\end{centering}

\caption{The spectra and physical quantities related to the tranmission coefficient
$T_{0,0}$ for a transversally symmetric short cavity: (a) the amplitude
of transmission coefficient $T_{0,0}$, (b) and (c) the amplitude
and phase (rad) of a component of $T_{0,0}$, \emph{i.e.}, $T_{0,0,\mu}$
versus $k$, (d) and (e) the real and imaginary parts of eigenvalue
$K_{\mu}$ versus $k$.}
\end{figure}

Figure 4(a) depicts the amplitude of the transmission coefficient
of the $0^{th}$ waveguide mode, $T_{0,0}$ , for a long cavity ($d=1$
m, $L_{x}=5$ m and $L_{y}=2$ m), as shown in Fig. 2(a). Within the
frequency range of interest ($k<3.5\ \mathrm{m^{-1}}$), the components
of $T_{0,0}$, \emph{i.e.}, $|T_{0,0,\mu}|$ are mainly due to the
first five eigen-modes ($\mu=(0,0),(1,0),(2,0),(3,0),(4,0)$) of the
open cavity. Figure 4(b) shows that $|T_{0,0,\mu}|$ are also characterized
by their resonance peaks. The comparison between Fig. 3(d) and Fig.
4(g) suggests that the local extrema in $|T_{0,0}|$ of the long cavity
are also due to the superposition of eigenmodes. For example, at $k=0.68\ \mathrm{m^{-1}}$,
the maximum of $|T_{0,0}|$ is not caused by the resonance peak of
$|T_{0,0,(1,0)}|$ alone, while $|T_{0,0,(0,0)}|$ and $|T_{0,0,(2,0)}|$
also have non-negligible contributions to the transmission coefficient.
However, the latter two terms cancelled each other leaving the (1,0)
eigenmode of the cavity as the sole contributor to $T_{0,0}$ at frequencies
around $k=0.68\ \mathrm{m^{-1}}$. A similar analysis can be made
for the other $|T_{0,0}|$ peaks, except for the first one at $k=0\ \mathrm{m^{-1}}$,
which is a case of resonance transmission (to be discussed in Sec.
3.3). The dips in $|T_{0,0}|$, on the other hand, are due to the
destructive interference of the eigenmodes.

\begin{center}
\begin{figure}
\begin{centering}
\includegraphics[scale=1]{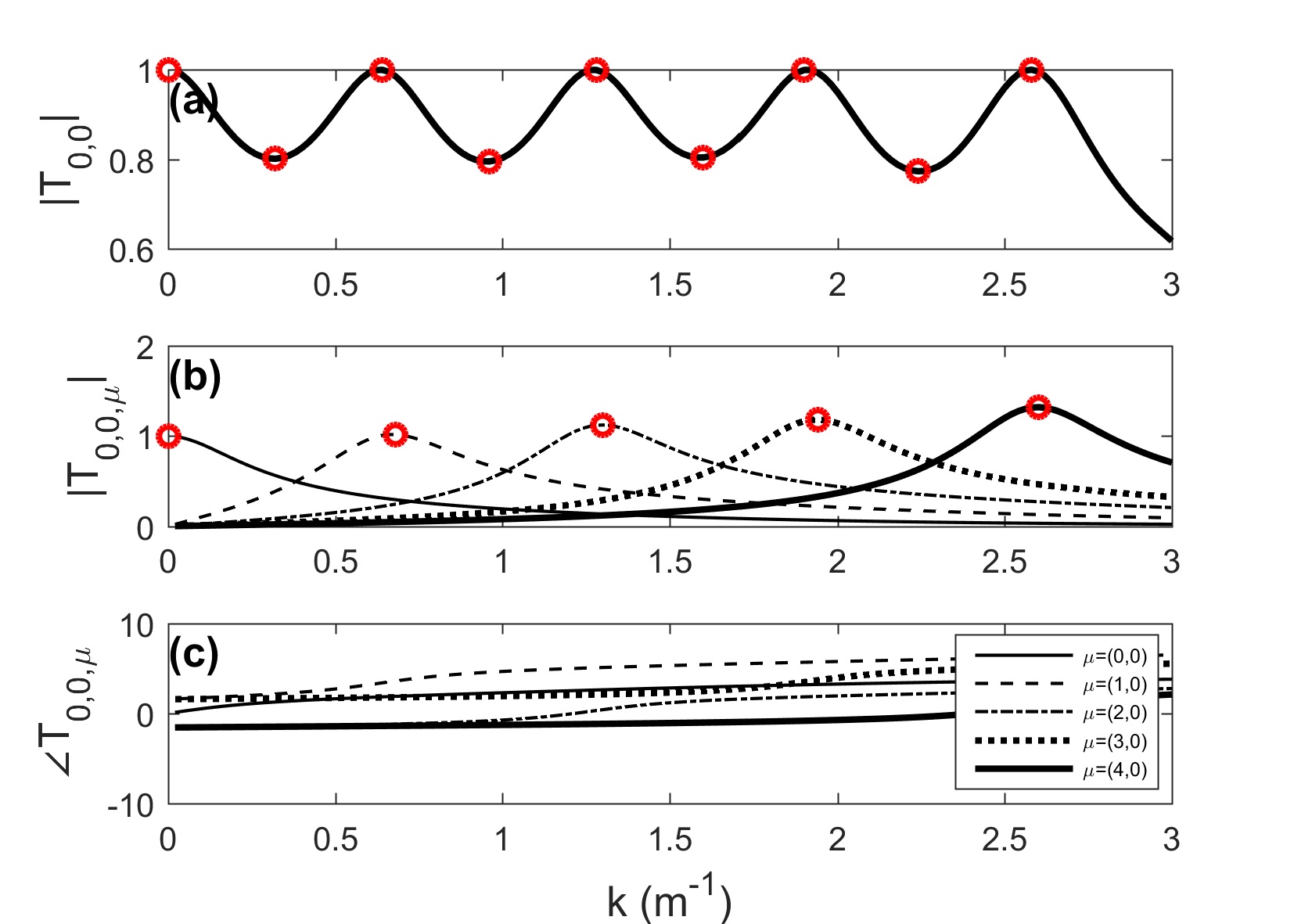}
\par\end{centering}

\caption{The spectra and physical quantities related to the transmission coefficient
$T_{0,0}$ for a long chamber: (a) the amplitude of transmission coefficient
$T_{0,0}$, (b) and (c) the amplitude and phase (rad) of five components
of $T_{0,0}$. }
\end{figure}

\par\end{center}

\subsection{Fano resonance}

The eigenmodes responsible for the transmission coefficients in Figs.
3 and 4 have relatively large imaginary parts. The open cavity also
has another type of eigenmodes, known as quasi-trapped modes, whose
eigenvalues have very small imaginary part and hence result in very
narrow resonance peaks. Although the involvement of quasi-trapped
modes in producing the sharp asymmetric Fano resonance is well known
\cite{hein2010fano,hein2012trapped,xiong2016fano}, the theoretical
analysis in Sec. 2. will be used to explain how these quasi-trapped
modes are involved in producing the Fano resonances.

\begin{figure}
\begin{centering}
\includegraphics[scale=1]{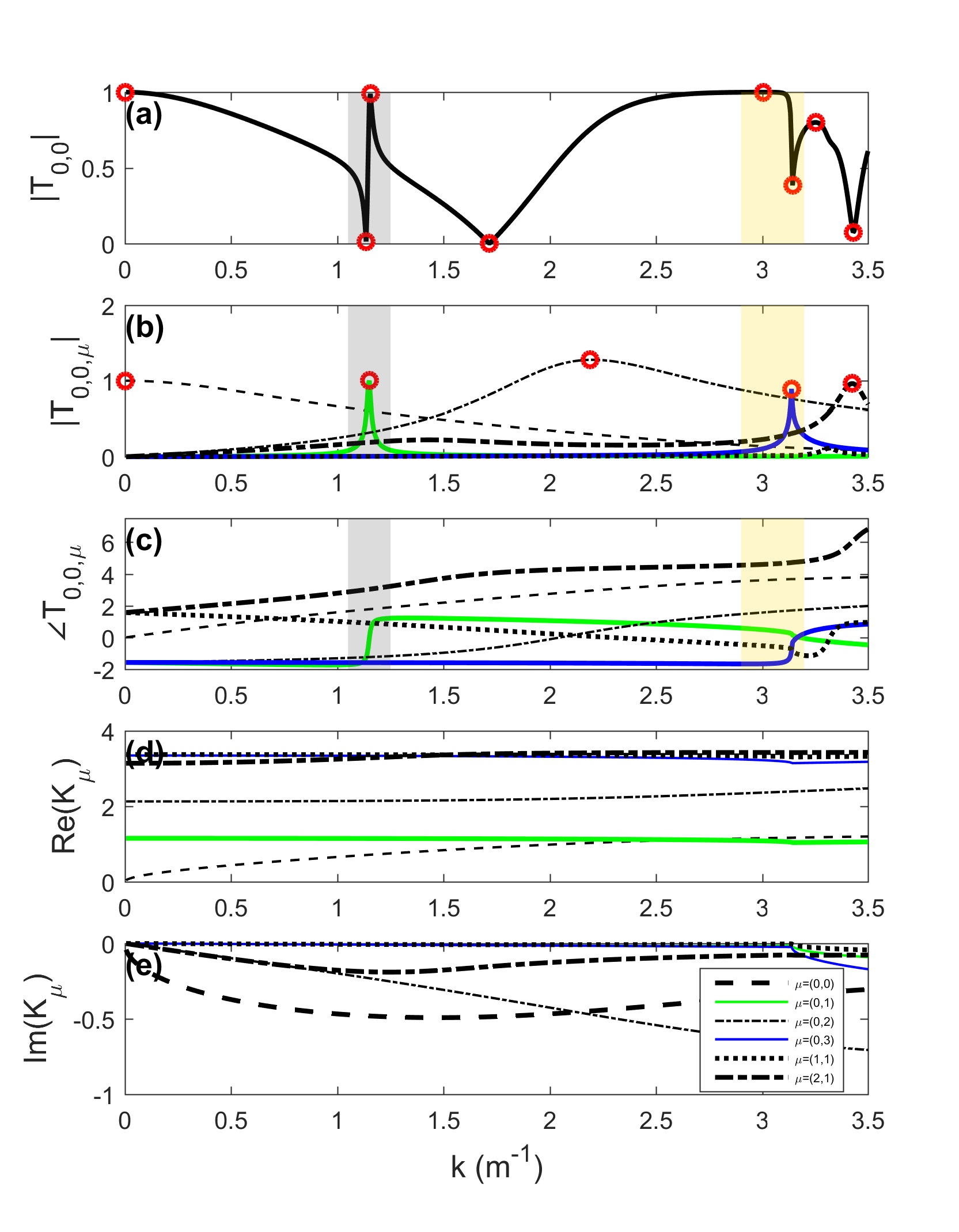}
\par\end{centering}

\caption{\label{fig:spectra-aysm-00}Spectra and physical quantities related
to transmission coefficient $T_{0,0}$ for a transversally asymmetric
short duct: (a) The amplitude of $T_{0,0}$, (b) and (c) the amplitude
and phase (rad) of $T_{0,0,\mu}$, and (d) and (e) the real and imaginar
parts of $K_{\mu}$. }
\end{figure}

Consider again the short cavity with parameters of $d=1$ m, $L_{x}=1$
m, $L_{y}=3$ m, and $\delta=0.1$ m. For this case the quasi-trapped
modes emerge as an offset is introduced to make the cavity asymmetric
about the central axis \cite{hein2012trapped}. Figure 5(a) presents
the amplitude of $|T_{0,0}|$ versus $k$. Compared with Fig. 3(a)
, it is found that extra peaks and dips emerge around $k=1.15\ \mathrm{m^{-1}}$
and $k=3.14\ \mathrm{m^{-1}}$ in Fig 5(a), showing typical Fano resonances.
To explain the formation of the Fano resonances, the contribution
to $T_{0,0}$ of each individual mode is plotted in Fig. 5(b). Due
to the broken transversal symmetry, the original fully localized modes
(0,1) and (0,3) modes become quasi trapped modes with $Re(K_{(0,1)})\sim1.15\ \mathrm{m^{-1}}$
and $Re(K_{(0,3)})\sim3.14\ \mathrm{m^{-1}}$, now participate in
the coupling with the waveguide waves. As shown in Figs 5(b) and (c),
that $K_{(0,1)}$ and $K_{(0,3)}$ now have very small imaginary parts.
In Fig. \ref{fig:spectra-aysm-00}(b), $|T_{0,0,(0,1)}|$ and $|T_{0,0,(0,3)}|$
display sharp resonance behaviors; the corresponding sharp phase changes
are also observed in Fig. \ref{fig:spectra-aysm-00}(c) at the peak
frequencies.

An interpretation of this phenomenon is straightforward following
the wave superposition expression in Eq. (\ref{eq:new scattering formula}).
Taking the first Fano resonance ( $k\approx1.15\mathrm{\ m^{-1}}$)
as an example, in the vicinity of $k=1.15\ \mathrm{m^{-1}}$, the
quasi-trapped mode (0,1) demonstrates a resonance response, and hence
$T_{0,0,(0,1)}$ has a sharp phase change, while the responses of
the other modes vary slowly in terms of amplitude and phase. Therefore,
below and above the resonance frequency of $T_{0,0,(0,1)}$, constructive
and destructive interferences between $T_{0,0,(0,1)}$ and components
of other eigenmodes give rise to a local maximum and minimum, respectively,
of the $T_{0,0}$. The contribution by non-resonance modes can be
approximated by a complex constant. As a result, the Fano resonance
can be characterized as the superposition of a resonantly excited
eigenmode $\mu$ (here, $\mu=(0,1)$) and background non-resonance
response $c_{1}$, \emph{i.e.},

\begin{eqnarray}
T_{0,0} & = & T_{0,0,\mu}+\sum_{\mu'\neq\mu}T_{0,0,\mu'}\nonumber \\
 & \approx & T_{0,0,\mu}+c_{1}\nonumber \\
 & \approx & \frac{c_{2}}{\mathcal{K}_{\mu}^{2}-k^{2}}+c_{1}\label{eq:appox Fano formula}
\end{eqnarray}
where in the third row, the numerator of the first term, originally
expressed as $\kappa_{R,0}H_{\mu,\nu,L,0}H_{R,0,\mu,\nu}$ in Eq.
(\ref{eq:new scattering formula}), is approximated by a constant
$c_{2}$, and $K_{\mu}$ in the denominator by a constant ${\cal K}_{\mu}$.
The approximations are valid for the narrow bandwidth in which Fano
resonance occurs, as the approximated terms vary slowly with frequency.
The frequency-dependent $K_{\mu}$ can be taken as a constant $\mathcal{K}_{\mu}$,
as $K_{\mu}$ is almost a constant within the frequency band. 

On the other hand, it is well known that the standard Fano formula
\cite{miroshnichenko2010fano} in terms of sound power transmission,
in \emph{the vicinity} of $Re(K_{(0,1)})$ can be written as:

\begin{equation}
T_{standard}(k)=\sigma\left(q+\frac{k-Re({\cal K}_{(0,1)})}{Im({\cal K}_{(0,1)})}\right)^{2}/\left[1+\left(\frac{k-Re({\cal K}_{(0,1)})}{Im({\cal K}_{(0,1)})}\right)^{2}\right],\label{eq:standard fano formula}
\end{equation}
where $(k-Re({\cal K}_{(0,1)}))/Im({\cal K}_{(0,1)})$ is the reduced
resonance frequency, $q$ measures the asymmetry of the resonance
shape, and $\sigma$ is the normalization constant.

\begin{figure}
\begin{centering}
\includegraphics{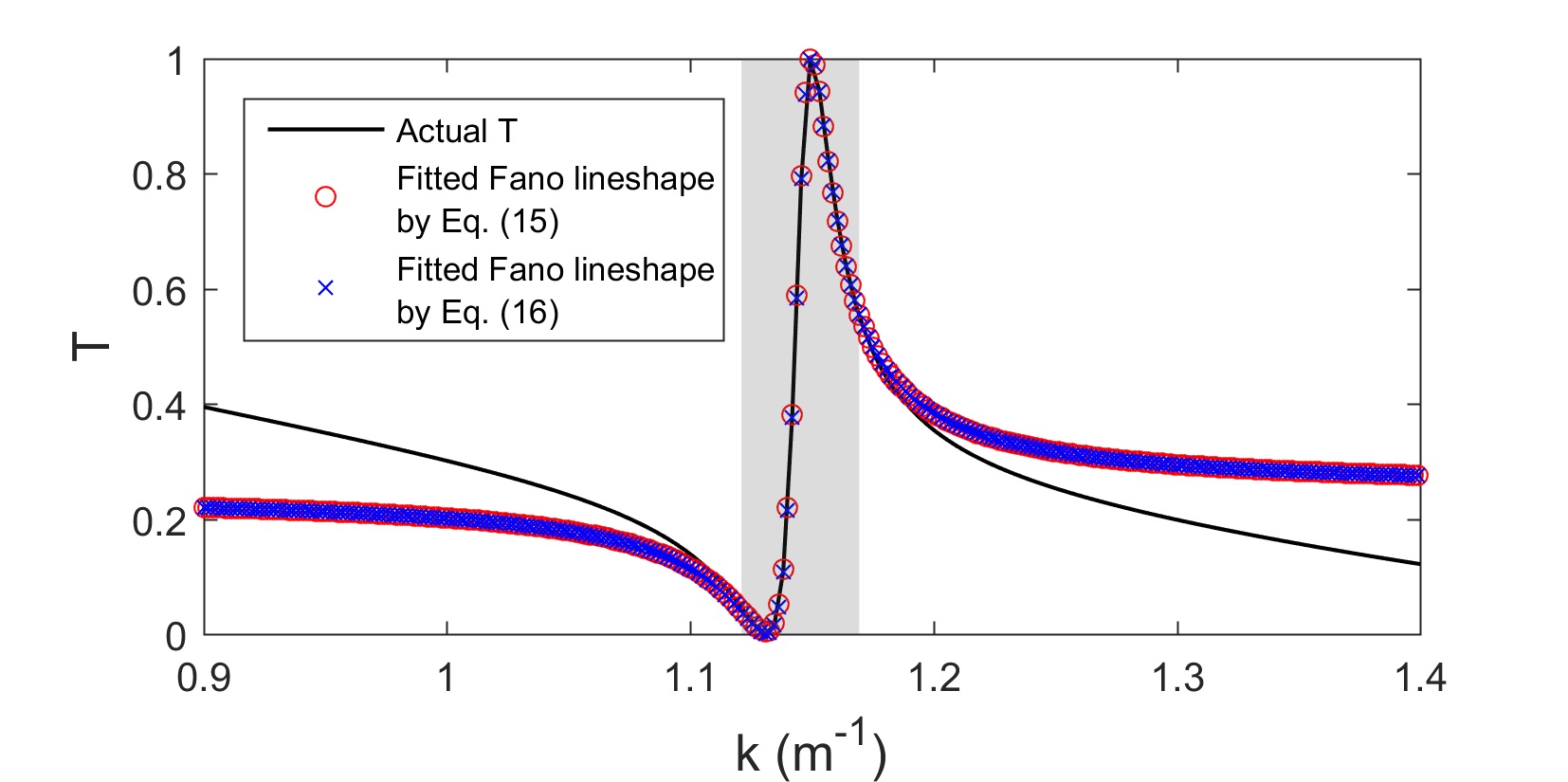}
\par\end{centering}

\caption{\label{fig:curvefitting}The sound power transmission coefficient
$T$: (solid line) actual $T(k)$ calculated using Eq. (\ref{eq:Tdecomposition}),
(circles) fitted curve using Eq. (\ref{eq:appox Fano formula}) and
(crosses) fitted curve using Eq. (\ref{eq:standard fano formula}). }
\end{figure}

A way to verify Eq. (\ref{eq:appox Fano formula}) is to fit the $T_{0,0}(k)$
to the actual sound power transmission curve to determine the value
of ${\cal K}_{\mu}$ and check if it agrees with $K_{\mu}$ by eigensolution
(Eq. (\ref{eq:EVP2})) within the narrow band. It is shown in Fig.
\ref{fig:curvefitting} the original and fitted curve of $T(k)=|T_{0,0}|^{2}$
using Eq. (\ref{eq:appox Fano formula}) in Fig. \ref{fig:curvefitting}.,
Excellent agreement is observed between the actual sound power transmission
curve and the fitted curve using Eq. (\ref{eq:appox Fano formula}),
while the fitted parameter ${\cal K}_{(0,1)}$ is also very close
to the frequency-dependent eigen-value $K_{(0,1)}$ at the resonance
frequency for $k=1.145\ \mathrm{m^{-1}}$ of the (0,1) eigenmode,
as is shown in Table. \ref{tab:Eigen-values-obtained}. The results
of curve fitting indicate that Eq. (\ref{eq:appox Fano formula})
indeed serves as a good description of the Fano resonance caused by
the quasi trapped mode (0,1).

A further observation, according to Fig. (\ref{fig:curvefitting})
and Table. \ref{tab:Eigen-values-obtained}, is that the fitted curve
using Eq. (\ref{eq:appox Fano formula}) agrees well with that using
the standard Fano formula in Eq. (\ref{eq:standard fano formula}).
It is interesting that the two fitted curves exhibit excellent agreement
not only within the narrow band where Fano resonance occurs, but at
other frequencies as well. It is an indicator that Eq. (\ref{eq:appox Fano formula})
is actually equivalent to the standard Fano formula - they cannot
agree globally unless they are describing the same function of $k$,
or more exlicitly, they agree globally because they are equivalent
and are describing the same function of $k$.

\begin{table}[H]
\begin{centering}
\begin{tabular*}{12.5cm}{@{\extracolsep{\fill}}>{\centering}p{2cm}>{\centering}p{2cm}>{\centering}p{2cm}>{\centering}p{5.5cm}}
\hline 
Methods & $Re(K_{(0,1)})$ or $Re({\cal K}_{(0,1)})$ & $Im(K_{(0,1)})$ or $Im({\cal K}_{(0,1)})$ & \multicolumn{1}{>{\centering}p{5.5cm}}{Parameters}\tabularnewline
\hline 
EVP ($k=1.145\ \mathrm{m^{-1}}$) & 1.146 & -0.0848 & \tabularnewline
\hline 
Curve fitting I & 1.145 & -0.0836 & $\left|c_{2}\right|$= 0.0174, $c_{1}/c_{2}=-26.31+11.73i$ \tabularnewline
\hline 
Curve fitting II & 1.145 & -0.0839 & $q=1.744$, $\sigma=0.2478$\tabularnewline
\hline 
\end{tabular*}
\par\end{centering}

\caption{Eigenvalues obtained by (1) solving the EVP directly and (2) and (3)
curve fitting using Eq. (\ref{eq:appox Fano formula}) and (\ref{eq:standard fano formula}),
respectively.\label{tab:Eigen-values-obtained}}
\end{table}

\subsection{Resonance transmission}

The first local maximum in $|T_{0,0}|$ in Figs. 3(a), 4(a) and 5(a),
at $k=0\ \mathrm{m^{-1}}$, is contributed by the resonance peak of
$|T_{0,0,(0,0)}|$ only , while the components of all other eigenmodes
are negligibly small. In other words, the transmission peak at $k=0\ \mathrm{m}^{-1}$
in $|T_{0,0}|$ is dictated by the resonance of $(0,0)$ mode alone,
and hence is called ``resonance transmission'' to distinguish it
from the cases discussed in Sec. 3.1 and 3.2, where destructive and
constructive superposition of eigenmodes led to local extrema in the
scattering coefficients. The condition for the occurrence of resonance
transmission is very strict in that it requires the scattering to
be dominated by individual modes. In some cases, these conditions
are fulfilled such that resonance transmission can be observed. It
is considered the scattering coefficient $T_{1,0}$ for the same asymmetrical
short cavity discussed in Sec 3.2. Figures 7(a) and (b) present respectively
the amplitude of $T_{1,0}$ and the modal components $T_{1,0,\mu}$,
and the comparison of these indicates resonance transmission at $k=1.15\ \mathrm{m^{-1}}$
and $k=3.14\ \mathrm{m^{-1}}$. 

\begin{figure}
\begin{centering}
\includegraphics[scale=1]{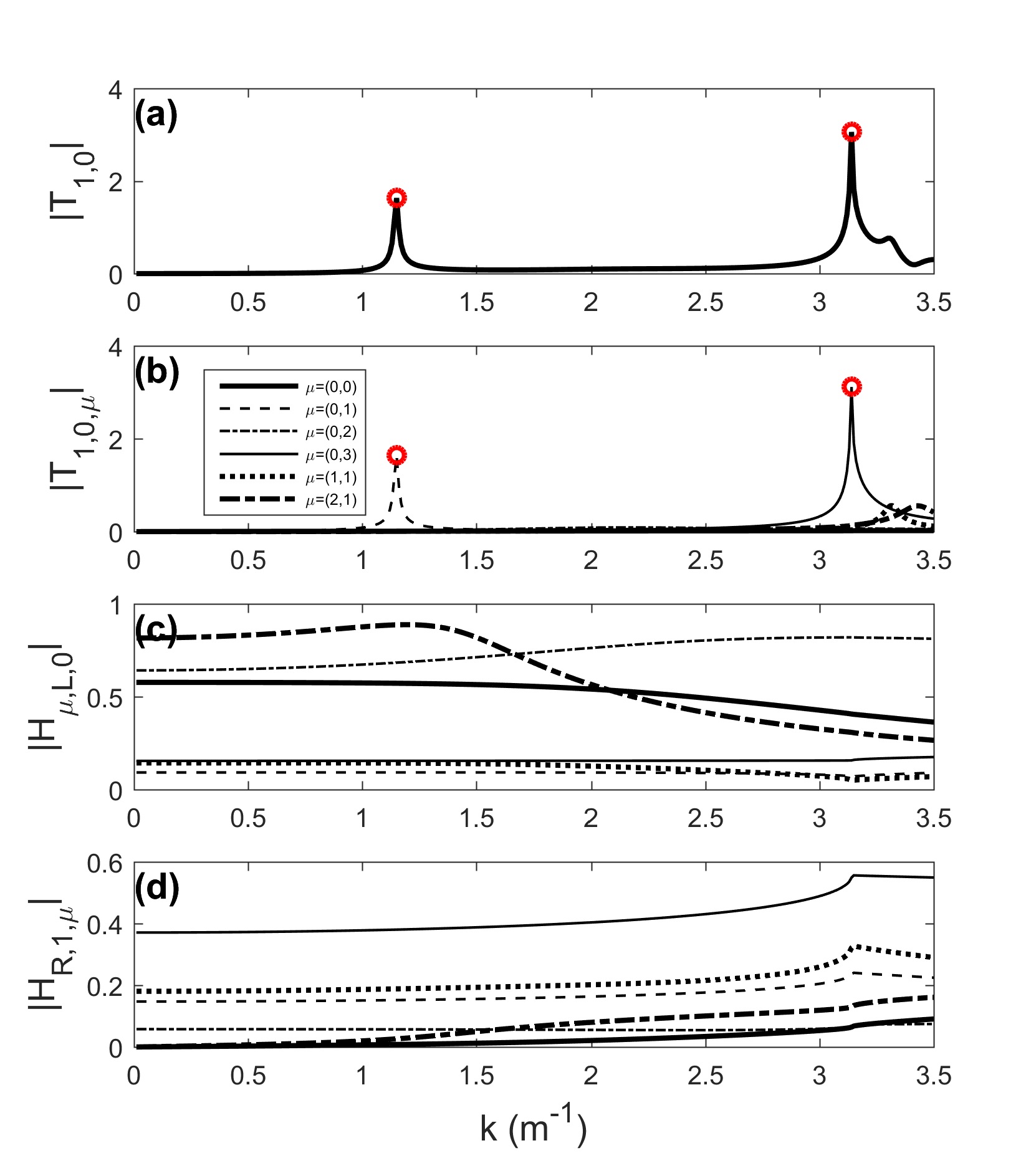}
\par\end{centering}

\caption{\label{fig:spectrum-asym-10}The amplitude of (a) $T_{1,0}$, (b)$T_{1,0,\mu}$,
(c) $H_{\mu,L,0}$, and (d) $H_{R,1,\mu}$, for a short cavity.}
\end{figure}

At the first resonance transmission ( $k=1.15\ \mathrm{m^{-1}}$ ),
$T_{1,0}$ is almost dominated by $T_{1,0,(0,1)}$, the modal component
by ($0,1$) eigen-mode. Figures. \ref{fig:spectrum-asym-10}(d) and
(e) demonstrate that the modal coupling factors $H_{\mu,L,0}$ and
$H_{R,1,\mu}$ exert a crucial influence, as generally only when both
of them take relatively large values can $T_{1,0,\mu}$ play an important
role in $T_{1,0}$. For example, $T_{1,0,(0,0)}$ correponds to a
small $|H_{R,1,(0,0)}|$ and a large $|H_{(0,0),L,0}|$, leading to
an ignorable transmitted wave $b_{R,1}$ owing to the poor coupling
between the $(0,0)$ eigenmodes and the transmitted first order duct
modes, despite the fact that the $(0,0)$ eigenmode is sufficiently
excited by the incident plane wave $a_{L,0}$. A similar reasoning
applies to $T_{1,0,(0,2)}$, where the (0,2) eigenmode has a negligible
contribution to $T_{1,0}$ owing to the small $|H_{R,1,(0,2)}|$,
\emph{i.e.}, the poor coupling between the (0,2) modes and the transmitted
first order duct mode.

The above observation, however, does not hold for the squasi trapped
mode (0,1). Although $|H_{(0,1),L,0}|$ is small (owing to weak coupling
between the $(0,1)$ eigenmode and the incident plane wave $a_{L,0}$)
, the component of $T_{1,0}$ from the $(0,1)$ mode is still large.
It is reasoned that the eigenvalue $K_{(0,1)}$ of the quasi trapped
mode $(0,1)$ is almost real, giving $1/(K_{(0,1)}^{2}-k^{2})$ in
Eq. (\ref{eq:MPT}), and hence the $T_{1,0,(0,1)}$ term very large
value when $k$ is close to $Re(K_{(0,1)})$. Therefore, in such a
situation, only the (0,1) eigenmodes contributes significantly to
the $T_{1,0}$ at $k=1.15\ \mathrm{m^{-1}}$, where $|T_{1,0}|$ achieves
its local maxima. In the same way, it is possible to show that the
resonance transmission close to $k=3.14\ \mathrm{m^{-1}}$ is dominated
by (0,3) mode.

It is noted that, in Fig. 7(a) $|T_{1,0}|$ is sometimes greater than
1. For the peak around $k=1.15\ \mathrm{m^{-1}}$, $|T_{1,0}|$ takes
a value of around 1.8, however, it does not correspond to any transmitted
sound power as the first-order mode is evanescent and carrying zero
net energy flux when $k<\pi\ m^{-1}$. At frequencies around $k=3.14\ \mathrm{m^{-1}}$,
the axial wavenumber is very small, meaning that the transmitted sound
power carried by the first order duct mode is actually less than that
carried by the incident zeroth-order duct mode, even when $|T_{1,0}|$
is greater than unity.


\section{Further discussion about Eq. (\ref{eq:EVP2})}

In Eq. (\ref{eq:new scattering formula}), the bi-orthogonal eigenmodes
defined by Eq. (\ref{eq:EVP2}) are used to decompose the scattering
coefficients, and have been employed throughout this paper. They are
\emph{frequency-dependent}. It is noted that Eq. (\ref{eq:acousticscatteringformula})
can also be used to define another type of EVP, which gives rise to
the \emph{frequency-independent} eigenmodes, satisfying,:

\begin{equation}
\boldsymbol{D}(\widetilde{k}_{\mu})\widetilde{\boldsymbol{g}}_{\mu}=\tilde{k}_{\mu}^{2}\widetilde{\boldsymbol{g}}_{\mu},\label{eq:EVP I}
\end{equation}
where $\boldsymbol{D}$ is defined by Eq. (\ref{eq:effectivehamiltonian})
but is a function of eigenvalue $\widetilde{k}_{\mu}$. 

The solution of the frequency-independent EVP requires 

\begin{equation}
\det\left(\boldsymbol{D}(\widetilde{k}_{\mu})-\tilde{k}_{\mu}^{2}\boldsymbol{I}\right)=0.
\end{equation}
As Eq. (\ref{eq:acousticscatteringformula}) includes the inverse
of $\left(\boldsymbol{D}-k^{2}\boldsymbol{I}\right)$, it suggests
that eigenvalues of Eq. (\ref{eq:EVP I}) are exactly the poles of
$s_{n,n',p,p'}$ (and, in general, the scattering matrix). In general,
$\widetilde{k}_{\mu}$ takes complex value, and is known as the complex
resonance frequency of the scatterer, with $Re(\widetilde{k}_{\mu})$
being the resonance frequency and $Im(\widetilde{k}_{\mu})$ characterizing
the decay of the mode. For some special cases, $\widetilde{k}_{\mu}$
takes a real value, corresponding to localized mode with no radiation
loss , which is usually referred to as a trapped mode or bound state
in the continuum \cite{lyapina2015bound,linton2007embeddedtrappedmode,xiong2016fano}. 

As defined, the frequency-dependent EVP yields the bi-orthogonal eigenmodes
for the forced response of the scatterer, whilst the frequency-independent
EVP produces the modal characteristics of the free vibration of the
system. As the response of the scatterer to an incident wave is a
forced vibration problem, the use of the frequency-dependent eigenmodes
is appropriate. 

However, in some previous research, frequency-independent eigenmodes
were sometimes adopted to deal with the forced response of the system
\cite{hein2012trapped,pelat2009use,yang2013mechanism,yu2016acoustic}.
Owing to the nonlinearity of Eq. (\ref{eq:EVP I}), and hence the
absence of orthogonality and completeness of $\left\{ \widetilde{\boldsymbol{g}}_{\mu}\right\} $
as a complete set, the modal expansion of the sound field by the frequency-independent
bases $\left\{ \widetilde{\boldsymbol{g}}_{\mu}\right\} $ lacks justification.

A typical issue encountered when the frequency-independent eigenmodes
are used for a forced response problem can be found in Ref. \cite{yu2016acoustic},
where some peaks/dips in transmission loss (TL) at continuous right-angled
corners were attributed to even/odd distributed frequency-independent
eigenmodes. An analysis following the procedures in the present paper,
however, shows that the extrema in the TL are, as expected, the result
of constructive/destructive interference of frequency-dependent eigenmodes,
rather than the product of single eigenmodes. Examples of this issue
can also be found in the modelling of sound propagation in street
canyons using non-orthogonal frequency-independent eigenmodes \cite{pelat2009use}
and in the expansion of the sound field in and outside a parallel sound
barrier by frequency-independent eigenmodes \cite{yang2013mechanism}. 

However, the frequency-independent eigensolutions may be used for
the forced response of the scatterer in some special cases when (quasi-)
trapped modes are involved. Hein \emph{et. al.}, \cite{hein2012trapped},
employed a finite element solver to tackle complex resonance for the
investigation of quasi-trapped modes and corresponding Fano-resonance
in duct-cavity system. It is reasonable to do so because within the
narrow band of quasi-trapped modes, Eq. (\ref{eq:appox Fano formula})
can be further approximated to (readers may refer to the discussion
in Sec. 3.2)

\begin{figure}
\centering{}\includegraphics[scale=1]{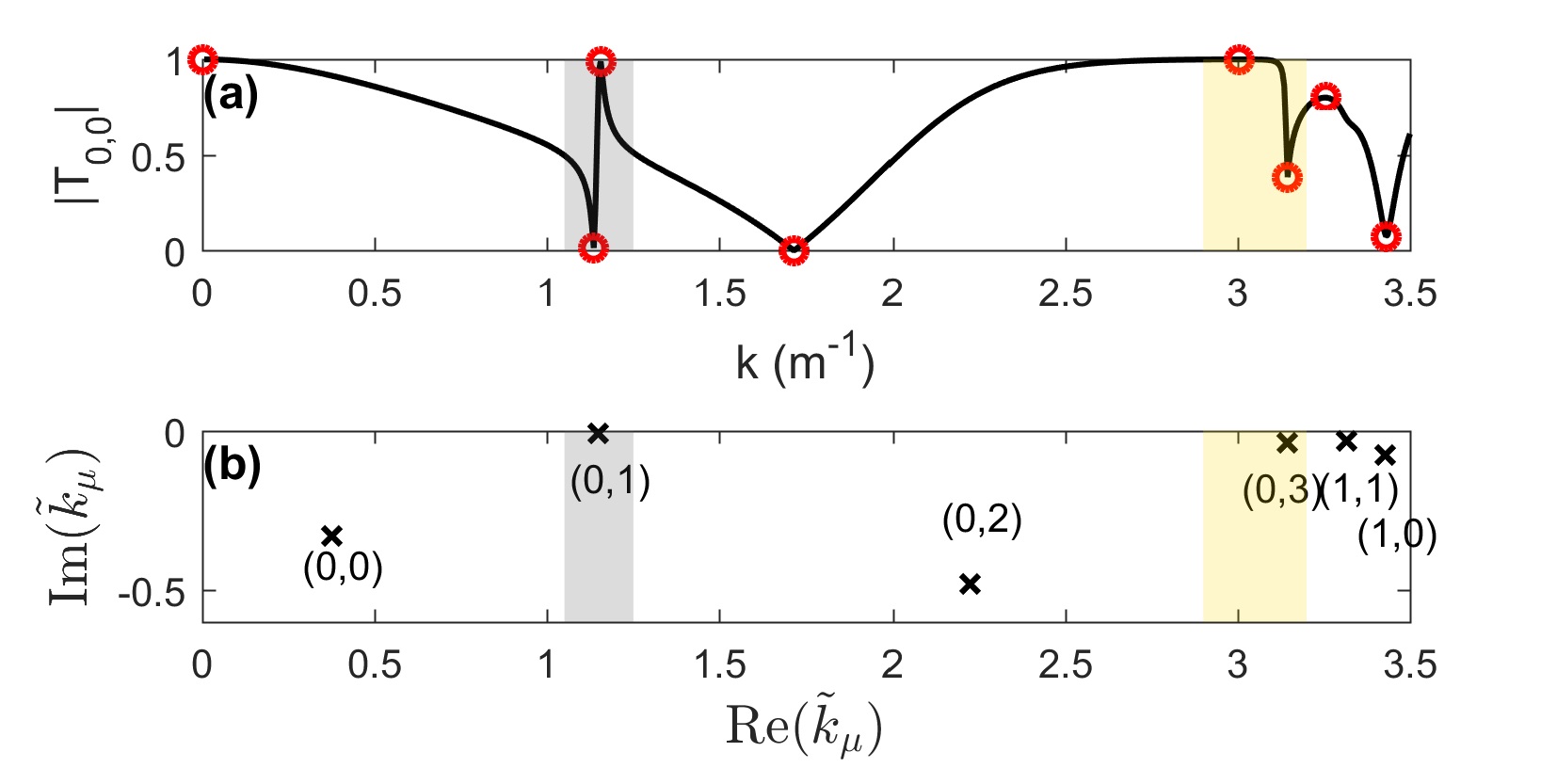}\caption{\label{fig:The-force-response}The force response of the asymmetric
short cavity considered in Sec. 3.2: (a) amplitude of $T_{0,0}$ and
(b) the frequency-independent eigenvalues $\widetilde{k}_{\mu}$
of the open cavity. }
\end{figure}

\begin{equation}
T_{0,0}\approx\frac{c_{2}}{{\cal K}_{\mu}^{2}-k^{2}}+c_{1}\approx\frac{c_{2}}{\widetilde{k}_{\mu}^{2}-k^{2}}+c_{1},\label{eq:furthe approx fano}
\end{equation}
by noting the fact from Fig. \ref{fig:The-force-response} that frequency-dependent
eigenvalue $K_{\mu}$ can be well approximated by frequency-independent
eigenvalue $\widetilde{k}_{\mu}$. Hence, Eq. (\ref{eq:furthe approx fano})
suggests that $\widetilde{k}_{\mu}$, the frequency-independent eigensolution
is able to characterize the frequency dependency of scattering coefficient
in such a special case. 


\section{Conclusions}

In this paper, the scattering coefficients of open cavities in an
acoustical waveguide were investigated via analysis of eigenmodes.
By utilizing the bi-orthogonal frequency-dependent eigenmodes, the
scattering coefficients can be explicitly expressed as the superposition
of eigenmodes, allowing for a study of the roles that eigenmodes play
in acoustic scattering.

It was shown that the local extrema in the scattering coefficients
generally result from the constructive and destructive interference
of the resonance and non-resonance modes. In particular, an approximation
equation was derived to describe the sharp asymmetric line-shape of
scattering coefficients within the narrow-band of Fano resonance induced
by quasi-trapped modes. The equivalence of the approximation equation
and the standard Fano formula was demonstrated. In addition, it was
demonstrated that, in some special cases, the sound scattering coefficient
is dominated by single modes, giving rise to resonance transmission 

The theoretical analysis was limited to two-dimensional acoustic resonators,
and the numerical study was restricted to resonators with simple rectangular
cavities. However, it is straightforward to extend the analysis to
three-dimensional irregular resonators, where more complicated phenomena
may be encountered in the analysis of the frequency-dependent features
of of scattering coefficients.


\section*{Acknowledgements}

The financial support from the Australian Research Council (ARC LP)
is gratefully acknowledged. The first author is also grateful for
the sponsorship from the China Scholarship Council.


\appendix
\section{Bi-orthogonal relation in Eq. (\ref{eq:bi-orthogonalrelation})}

Let $\lambda_{\mu}$ and $\boldsymbol{V}_{\mu}$ be the eigenvalue
and eigenvector, respectively, of the non-Hermitian matrix $\boldsymbol{\Gamma}$
such that

\begin{equation}
\boldsymbol{\Gamma}\boldsymbol{V}_{\mu}=\lambda_{\mu}\boldsymbol{V}_{\mu},\label{eq:EVPofGamma}
\end{equation}
where subscript $\mu$ is for labelling and $\lambda_{\mu}$ generally
take a complex value.

Similarly, let $\zeta_{\mu}$ and $\boldsymbol{W}_{\mu}$ be the eigenvalue
and eigenvector of the adjoint matrix of $\boldsymbol{\Gamma}$, \emph{i.e.},

\begin{equation}
\boldsymbol{\Gamma}^{\dagger}\boldsymbol{W}_{\mu}=\zeta_{\mu}\boldsymbol{W}_{\mu},\label{eq:EVPofAdjointOperator}
\end{equation}
where $\boldsymbol{\Gamma}^{\dagger}\neq\boldsymbol{\Gamma}$ since
$\boldsymbol{\Gamma}$ is non-Hermitian.

Taking the adjoint of Eq. (\ref{eq:EVPofAdjointOperator}) yields:

\begin{equation}
\boldsymbol{W}_{\mu}^{\dagger}\boldsymbol{\Gamma}=\overline{\zeta_{\mu}}\boldsymbol{W}_{\mu}^{\dagger}.\label{eq:adjointEqofEVP}
\end{equation}

By comparing Eq. (\ref{eq:EVPofGamma}) and Eq. (\ref{eq:adjointEqofEVP}),
one can derive the following correspondence from the uniqueness of
the eigenspectrum of $\boldsymbol{\Gamma}$:

\begin{equation}
\lambda_{\mu}=\overline{\zeta_{\mu}},\ all\ \mu,
\end{equation}

Right multiplying $\boldsymbol{V}_{\mu'}$ to the above equation and
left multiplying $\boldsymbol{W}_{\mu}^{\dagger}$ to Eq. (\ref{eq:EVPofGamma})
(where $\mu$ is replaced by $\mu'$) and subtracting the resulting
equations gives,

\begin{equation}
0=\left(\lambda_{\mu'}-\overline{\zeta_{\mu}}\right)\boldsymbol{W}_{\mu}^{\dagger}\boldsymbol{V}_{\mu'},
\end{equation}
such that the case $\lambda_{\mu'}\neq\overline{\zeta_{\mu}}$ holds
unless $\boldsymbol{W}_{\mu}^{\dagger}\boldsymbol{V}_{\mu'}=0$, \emph{i.e.}:

\begin{equation}
\boldsymbol{W}_{\mu}^{\dagger}\boldsymbol{V}_{\mu'}=\delta_{\mu',\mu}\boldsymbol{W}_{\mu}^{\dagger}\boldsymbol{V}_{\mu'}.\label{eq:bi-orthogonal-raw}
\end{equation}
With proper normalization, Eq. (\ref{eq:bi-orthogonal-raw}) gives
the normalized bi-orthogonal relation

\begin{equation}
\boldsymbol{W}_{\mu}^{\dagger}\boldsymbol{V}_{\mu'}=\delta_{\mu',\mu},\label{eq:bi-orthogonality-nonhermitian}
\end{equation}
 which relates the left and right eigenvectors of a non-Hermitian
matrix $\boldsymbol{\Gamma}$.

Furthermore, if $\boldsymbol{\Gamma}$ is symmetric, then the transpose
of Eq. (\ref{eq:EVPofGamma}) is

\begin{equation}
\boldsymbol{V}_{\mu}^{T}\boldsymbol{\Gamma}=\lambda_{\mu}\boldsymbol{V}_{\mu}^{T}.\label{eq:EVPofGamma-transpose}
\end{equation}
A comparison with Eq. (\ref{eq:EVPofAdjointOperator}) indicates $\boldsymbol{W}_{\mu}^{\dagger}=\boldsymbol{V}_{\mu}^{T}$
such that the bi-orthogonality by Eq. (\ref{eq:bi-orthogonality-nonhermitian})
is reduced to:

\begin{equation}
\boldsymbol{V}_{\mu'}^{T}\boldsymbol{V}_{\mu}=\delta_{\mu',\mu}.
\end{equation}


\section*{References}
\bibliographystyle{elsarticle-num}
\bibliography{references}

\end{document}